# Ab initio Molecular Dynamics Simulation of Threshold Displacement Energies and Defect Formation Energies in $Y_4Zr_3O_{12}$

Sruthi Mohan*, Gurpreet Kaur, C. David, B. K. Panigrahi, G. Amarendra

*Materials Physics Division, Material Science Group, Indira Gandhi Centre for Atomic Research, HBNI, Kalpakkam, India - 603102*

**sruthi@igcar.gov.in*

**Abstract**

Ab initio molecular dynamics simulations using VASP was employed to calculate threshold displacement energies and defect formation energies of $Y_4Zr_3O_{12}$ δ-phase, which is the most commonly found phase in newly developed Zr and Al-containing ODS steels. The Threshold displacement energy ($E_d$) values are determined to be 28 eV for $Zr_{3a}$ PKA along [111] direction, 40 eV for $Zr_{18f}$ atoms along [111] direction and 50 eV for Y recoils along [110] direction. Minimum $E_d$ values for O and O' atoms are 13 eV and 16 eV respectively. The displacement energies of anions are much smaller compared to cations, thus suggesting that anion disorder is more probable than cation disorder. All directions except the direction in which inherent structural vacancies are aligned, cations tend to occupy another cation site. The threshold displacement energies are larger than that of $Y_2Ti_2O_7$, the conventional precipitates in Ti containing ODS steels. Due to the partial occupancy of Y and Zr in the 18f position, the antisite formation energy is negligibly small, and it may help the structure to withstand more disorder upon irradiation. These results convey that Zr/Al ODS alloys, which have better corrosion resistance properties compared to the conventional Ti-ODS alloys, may also possess superior radiation resistance.

There is an increased demand for highly radiation tolerant and corrosion-resistant structural materials for core nuclear reactor applications. Ferritic steels, which have high incubation doses for void swelling, are better than austenitic stainless steels in nuclear core structural material applications[1]. However, ferritic steels are limited by its inferior high-temperature strength[2]. Oxide dispersion strengthened (ODS) ferritic steels, with high density of thermally stable and radiation-resistant dispersoids are observed to have high creep resistance along with radiation tolerance in reactor operating temperatures and doses[3]. Typical ODS steels containing Y–Ti–O dispersoids are found to be stable up to 150 dpa at 700°C[4]. Recently, there has been renewed interest in Al-containing ODS steels due to its enhanced corrosion resistance in supercritical water and lead bismuth eutectic environments[5]. However, upon addition of Al, fine (2-10 nm) titanium-rich Y- Ti – O complexes of Ti-ODS, which are mainly $Y_2Ti_2O_7$ or $YTiO_5$ gets replaced by coarse (20 – 100 nm) Y – Al – O complexes[6] and this deteriorates the ultimate tensile strength of the steel[7]. The addition of Zr refines the size of precipitates in Al-containing alloys to the desired size range of 2-10 nm [5, 8, 9]. The nanoparticles in Zr and Al containing ODS steels are identified to be rhombohedral δ- $Y_4Zr_3O_{12}$[9]. The radiation tolerance of ODS alloys vests on the radiation stability of the dispersoids[3].

$Y_4Zr_3O_{12}$ belongs to a class known as delta (δ) phase ($A_4B_3O_{12}$, $R\bar{3}$), which is an oxygen-deficient fluorite derivative structure($A_2B_2O_{8-δ}$ , $A^{3+}$ and $B^{4+}$ are rare-earth or transition metal species) like cubic pyrochlore ($A_2B_2O_7$, $Fd\bar{3}m$), disordered flourite ($A_2B_2O_7$, $Fm\bar{3}m$), and monoclinic pyrochlore ($A_2B_2O_7$, $P2_1$). These structures are also well known for their exceptional radiation tolerance and phase stability at elevated temperatures[10]. The properties of pyrochlores have been explored in depth over a wide compositional range and proposed for various applications, including electrolytes, sensors, containment for nuclear waste. Upon irradiation, most of the pyrochlores transform into disordered fluorite structure before they are fully amorphized[11]. The critical amorphization doses are less than 1 dpa for titanate pyrochlores[12]. Zirconate pyrochlores are more radiation-resistant than titanate pyrochlores. The experimentally determined amorphization dose of zirconate pyrochlore has been found to range from ~5.5 dpa (for $La_2Zr_2O_7$) to more than 100 dpa ($Gd_2Zr_2O_7$)[13]. The radiation stability of a pyrochlore depends on its degree of deviation from ideal fluorite structure[13].

However, there are only a few studies discussing the radiation stability of delta (δ) phase compounds. The rhombohedral to disordered fluorite transformation has been observed in $Sc_4Zr_3O_{12}$ upon swift heavy ion irradiation, which is carried out at room temperature(185 MeV $Au^{13+}$ ions, $1×10^{13}$ ions/cm², ~0.2 dpa)[14]. Such transformations have also been observed during low energy ion irradiation carried out at cryogenic temperature (300 keV $Kr^{2+}$ ions, $3×10^{16}$ ions/cm², 23 dpa ,100K)[15]. K. E. Sickafus *et al.* irradiated $Dy_4Zr_3O_{12}$, $Lu_4Zr_3O_{12}$ and $Sc_4Zr_3O_{12}$ with 300 keV $Kr^{2+}$ ions at 100K, and found that $Dy_4Zr_3O_{12}$ which was the most radiation-resistant, did not amorphize up to 55 dpa. Most importantly it was found the amorphization dose scales up with decrease in the order to disorder reaction pair energy[16]. Uranium containing δ-phase oxides $Y_6U_1O_{12}$, $Gd_6U_1O_{12}$, $Ho_6U_1O_{12}$, $Yb_6U_1O_{12}$, and $Lu_6U_1O_{12}$ showed resistance to amorphization up to 65 dpa of $Kr^{2+}$, $Ne^{2+}$ and $He^+$ at cryogenic temperature (~100K), but all of them showed partial or full ordered rhombohedral to disordered fluorite phase transformation[17]. Nevertheless, there is no report of ion irradiation-induced amorphization study in $Y_4Zr_3O_{12}$.

In order to produce a point defect by an energetic projectile, a sufficient amount of energy has to be transferred to a lattice atom. Threshold displacement energy (TDE) is the minimum kinetic energy required to displace an atom from its lattice site to create a stable defect[18]. This is the most fundamental quantity governing the creation and evolution of radiation cascade in a material[19] and thus a key parameter in Kinchin – Pease[20] and Norgett – Torrens – Robinson (NRT)[21] model to estimate the number of Frenkel pairs that an energetic primary knock-on atom (PKA) will create in a material. TDE is a critical factor for determining damage production profiles using Monte Carlo simulations and normalizing doses from various irradiation sources. Most experimental results in TDE are obtained by irradiating with energetic electrons[22, 23]. However, the short time and length scales of the radiation damage processes make these experiments challenging. Therefore, atomistic modeling is a promising tool for the understanding of the underlying mechanisms involved in primary defect production.

Classical molecular dynamics simulations have been used extensively to compute threshold displacement energies and evaluate recoil events in materials [24, 25]. However, the results of classical MD simulations are highly dependent on the quality of the empirical interatomic potential chosen, and it is likely to overestimate the value by ignoring the effects of partial charge transfer[24]. Recently, ab initio molecular dynamics (AIMD) has been employed as an alternative tool to study the low energy radiation response of metals[26], semiconductors[27-31], and ceramics. The advantage of AIMD is that since the forces are obtained by electronic structure calculations, the errors related to empirical potentials and partial charge transfer are avoided. Even though there are many reports on AIMD simulations of oxides[32-35], including pyrochlores[36-39], no such report exists in any δ-phase compound.

In this work, we have computed threshold displacement energies of Y, Zr, and O in $Y_4Zr_3O_{12}$ δ- phase compound along three main crystallographic directions using AIMD simulations and analyzed the defects formed and mechanisms involved in each case. The defect formation energies of $Y_4Zr_3O_{12}$ and the role of defects in rhombohedral to cubic transformation are discussed in comparison with cubic pyrochlores.

**Computational Details**

The calculations are done using the Vienna Ab-initio Simulation Package (VASP)[40] with pseudopotentials generated with the projected augmented wave approach (PAW)[41]. For exchange-correlation functional, the generalized gradient approximation as parameterized by Perdew, Burke, and Ermzerhof (PBE)[42] is used. For static calculations, a dense k point 5×5×5 Monkhorst-Pack[43] mesh is used, and the simulation cell of 152 atoms is fully relaxed (atom positions and cell volume) until the force on each atom is less than 0.001 eV/Å for both ideal structures and defect structures.

For AIMD calculations, plane wave energy cut off of 400 eV and automatic 1×1×1 mesh with their origin at the Γ point is used. An orthogonal supercell containing 456 atoms (shown in Fig.3.) was constructed from rhombohedral unitcell using the transformation matrix (-2,1,2; 2,1,2; 0,−2,2). The [100] direction in the original unitcell is equivalent to [322] direction in the transformed supercell. The simulations were carried out in NVE ensemble with periodic boundary conditions imposed along all the three axes. The supercell size is chosen in such a way that the image interactions are minimized. The supercells were first equilibrated at 300K for 3000 time-steps (300 fs). An atom ($Zr_{3a}$, $Y_{18f}$, $Zr_{18f}$ or $O_{18f}$) is chosen as the primary knock-on atom (PKA), and energy is given to it in three main crystallographic directions. After giving energy for a particular PKA in a particular crystallographic direction, the system is evolved in time with a variable time step method starting from 0.1 to 2 fs. The atom trajectories are traced for at least 2 ps. If the PKA is not returned to its original position after the completion of simulations (2ps), it is assumed to be a displaced atom. Here we study only primary damage creation and ignore the effects of long term defect annealing. Displacement energy is determined by ensuring that Frenkel pair or cation antisite defect is formed and is stable over sufficient time. All visualizations are done using VESTA[44].

**Results**

**Structure of $Y_4Zr_3O_{12}$**

The phase $Y_4Zr_3O_{12}$ was identified by Scott[45], Ray and Stubican[46] in the 1970s. Its structure is similar to $R_7O_{12}$ and $MR_6O_{12}$ (M being U,W or Mo and R a rare earth), belonging to $R\bar{3}$ space group with inversion triad along [111] direction of the rhombohedron[47]. The structure is shown in Fig.1. It has two ordered structural vacancies in 6c positions along its inversion triad. Oxygen gets occupied in two sets of 18f general positions. Both sets of oxygen form two octahedrons one being centered at body center (denoted as O) and the second (denoted as O') at 3a cation. Consequently, the cation in 3a position is co-ordinated to six oxygen atoms and that in 18f positions are coordinated to 7 oxygen atoms. The exact ordering of cations in 18f and 3a positions cannot be determined directly from X-ray or neutron diffraction data because Y and Zr have similar scattering factors. In 2001, Bogicevic et al.[48] searched for the lowest energy configurations of $Y_4Zr_3O_{12}$ using ab initio simulations and lattice algebra techniques and found that the six-fold coordinated cation site (3a) is occupied solely by Zr. By a combination of pair potential Monte Carlo simulations with DFT the ordered cation structures of a range of δ-phase compounds were computed and concluded that the ground state of $Y_4Zr_3O_{12}$ contains only one formula unit per unit cell. In this work, six different cation distributions are considered (Fig.2.) and compared the total energy for this 19 atom unitcell. The systems with Zr in 3a position have the lower energy compared to Y in 3a position and among the 18f cation positions, the two Zr occupying the diagonally opposite positions of the cation octahedron has lower energy compared to other combinations. The lowest energy structure is used for our further calculations. The configurations and corresponding formation energies per formula unit are summarized in the Fig.2.

**Relative stability of Pyrochlore, δ-phase and disordered fluorite in Y-Zr-O system**

The stacking arrangement of the metal ions in δ-phase remains the same as fluorite structure, except for the slight displacement of ions out of the fcc (111) planes. Similarly, a $Y_2Zr_2O_7$ pyrochlore structure (Space group: $Fd\bar{3}m$) can be derived from the cubic fluorite structure (space group $Fm\bar{3}m$) by removing one-eighth of the oxygen atoms. In $Y_2Zr_2O_7$, Y and Zr are in 16d and 16c sites. One set of oxygen is in 48f sites, surrounded by two Y and Zr ions each, and the second set of oxygen is in 8b site surrounded by four Y ions. The vacancy is at 8a site enclosed by four Zr ions.

The relative stability of pyrochlore, disordere-fluorite, and δ-phase structures containing Y, Zr, and O are compared by determining their formation energies with respect to $Y_2O_3$ and $ZrO_2$ using density functional theory. A similar study has been done before in $La_2Ce_2O_7$ to find the relative stability between pyrochlore and disordered fluorite structures[49]. Pyrochlore $Y_2Zr_2O_7$ is constructed from the $ZrO_2$ structure by replacing half of the $Zr^{4+}$ atoms using $Y^{3+}$ and removing one – eighth of oxygen with the same chemical environment.

Special quasirandom structure (SQS) is generally used to calculate the electronic and thermodynamic properties of random structures[50]. It is a periodic structure whose distinct correlation functions match with the ensemble average of the truly random structure. To mimic fully disordered cubic fluorite structure, an 88 atom SQS constructed by Jiang et al[51]. is used. The formation energies per atom of different structures are calculated as[49]:

$$E_{Y2Zr2O7} = \frac{E_{tot} - E_{Y2O3} - 2E_{ZrO2}}{N_f}$$

$$E_{Y4Zr3O12} = \frac{E_{tot} - 2E_{Y2O3} - 3E_{ZrO2}}{N_f}$$

$E_{Y2O3}$ and $E_{ZrO2}$ are the bulk energies of $Y_2O_3$ and $ZrO_2$ respectively and $N_f$ is the number of atoms the supercell contains.

The calculated formation energies are tabulated in Table. 2. The formation energies of both pyrochlores are almost similar irrespective of their oxygen vacancy environment, which is due to similar ionic radii of Y and Zr. Of the four systems considered, $Y_4Zr_3O_{12}$ has the most favored energy of formation and hence is energetically preferred to form (Fig.4.). The simulation results are in agreement with earlier studies based on the ionic radius ratio of cations in fluorite related structures. In the compositional dependence structural stability mapped by C.R. Stanek et al. [52], the experimentally observed $\delta$-phase series of compounds exist within a triangular compositional range drawn with vertices at $Sc^{3+}$-$Ti^{4+}$, $Sc^{3+}$ - $Zr^{4+}$ and $Yb^{3+}$ - $Zr^{4+}$. According to them, when the radius ratio($r_{A^{3+}}/r_{B^{4+}}$) is in between 1.21 and 1.42 $\delta$- phase is the most stable structure. The radius ratio $r_{Y^{3+}}/r_{Zr^{4+}}$ is 1.415 and is right on the middle of $\delta$-phase and pyrochlore.

**Displacement energies in $Y_4Zr_3O_{12}$**

For a chosen PKA atom and a particular crystallographic direction, the system is evolved for 2 ps with variable time-steps, as explained in the computational details section. The simulations are repeated with incremental PKA energy in steps of 1 eV. The minimum initial energy of PKA atom, which finally results in the displacement of PKA atom, is considered as the threshold displacement energy. The threshold displacement energies thus obtained and resulting defect configurations are summarized in Table.3. The number of anion defects is more compared to cation defects suggesting that anion disorder may be predominant under the radiation environment.

**Nature of defects**

Fig-5a shows the events occurring after imparting threshold energy along the [100] to Zr atom at 3a position. When the kinetic energy of 65eV is imparted to $Zr_{3a}$ atom along the [100] direction, it repulsively interacts with $Y_{18f}$ atom situated at 3.6 Å along this direction. The energy transferred to $Y_{18f}$ is sufficient to knock it out of its equilibrium position and form a cation antisite defect. Further, the path of $Y_{18f}$ recoil atom deviates from [100] direction to [631] direction, and it stabilizes at a bridge site between $Zr_{3a}$ and $Y_{18f}$ atoms, 2.9 Å away from its equilibrium site. An O' atom, which is initially bonded to secondary recoil atom ($Y_{18f}$), also gets displaced by 3.6 Å along [$\bar{5}$11] direction forming the O' Frenkel pair. The net defects produced are: $Y_{Zr}$ antisite, an O' Frenkel pair, a $Y_{18f}$ interstitial, and a vacancy at 3a site.

For a $Zr_{3a}$ PKA with $E_d$ of 60 eV along the [110] direction (Fig.5.b), the mechanism is similar to that in the [100] direction. It initially collides and replaces the Y atom in its path. The secondary recoil atom (SRA) moves further along the [110] direction and stabilizes in the bridge site between $Zr_{3a}$ and $Y_{18f}$ atoms. The distance moved by SRA is 3.5 Å. The O atom, which is initially bonded with PKA migrates by 3 Å along the same direction and knocks out an O' atom and occupies its position. The anion interstitial thus formed moves 2.8 Å away from its initial position and occupies a vacant 6c site. After these events, the following defects are produced: $Zr_{3a}$ vacancy, $Zr_Y$ antisite, $O_{O'}$ antisite, O' Frenkel pair, and a Y interstitial.

The lowest displacement energy of (28eV) Zr in 3a position is along [111] direction (Fig.5.c), the direction in which the structural anion vacancies are aligned. The only defect produced is a Zr interstitial - vacancy pair, separated by 4.7 Å, which stabilizes at body center. There is no cation antisite formation or anion migration. The two vacancies aligned in this direction make it easier to move for Zr atom and stabilize at the center of anion octahedron, where it has a similar anion environment as 3a site. Four out of six O atoms move 0.6 Å away from the equilibrium position to accommodate the Zr atom. The Zr forms new bonds with neighboring O atoms, and the average Zr-O distance is ~2.3 Å.

When displacement energy of 65 eV is imparted to the $Zr_{18f}$ along [100] direction (Fig.5.d): the PKA repulsively interacts with two Y atoms in the direction and changes its path along [23$\bar{1}$] direction, knocks off $Zr_{18f}$ atom in that direction, situated at the farthest point of cation octahedron forming a Zr interstitial. The secondary recoil atom gets moved to 2.4 Å away from its site. Two O' atoms which are in the direction of motion of the PKA are displaced by 2.3 Å and 2.8 Å respectively producing Frenkel pairs.

For $Zr_{18f}$ PKA along [110] direction (Fig.5.e), the threshold displacement energy is determined to be 69 eV. The PKA moves through the center of cation octahedron, collides with another $Zr_{18f}$ atom and replaces it. The recoiled Zr migrates 3.4 Å from its equilibrium position and occupies the vacant 6c site. O which was initially bound to PKA moves 3.4 Å along [110] direction and replaces O atom. The displaced O atom continues moving in the same direction eventually displacing an O' atom and occupies its position. The displaced O' atom moves 2.8 Å to form an interstitial.

$Zr_{18f}$ atom has the minimum threshold displacement energy of 40 eV along [111] direction (Fig.5.f). The $Zr_{18f}$ PKA moves along [111] direction and stabilizes as an interstitial at the bridge site between $Zr_{3a}$ and Y atom of the next cell, 5.12 Å away from its original position. The PKA also knocks out an O' atom situated in [111] direction. The knocked-out oxygen eventually moves along [111] direction, replacing another O' atom on its way. The recoiled O' atom migrates along [$\bar{3}\bar{2}$4] direction and eventually occupies one of the vacant 6c site (structural vacancy). The distance between the first O' atom displaced by PKA and the now occupied 6c site 3.3 Å. Another O' atom initially bonded to the PKA moves 2.8 Å from its original site and stabilizes in the 6c site of the nearby cell.

A similar approach has been employed for Y recoils. For Y PKA along [100] direction (Fig.5.g), the calculated threshold displacement energy is 65 eV. The Y PKA initially collides with another Y atom in [100] direction and occupies its site. The SRA moves further along [100] direction to form a stable interstitial at the face of the rhombohedral structure, at a distance of 3.2 Å. During this process, O atom which was initially bound to PKA moves along [3$\bar{1}$1] by 3.9 Å towards 6c site in the next cell. Another O' atom which was initially bonded with PKA moves by 2.6 Å along [100] direction to form a Frenkel pair.

For Y recoil along [110] direction(Fig.5.h), the threshold displacement energy is determined to be 50 eV. The Y PKA moves along [110] direction, eventually knocking of $Zr_{3a}$ atom at a distance of 5.06 Å in its path and occupying its position to form an antisite. The recoiled Zr atom form a stable interstitial at a distance 3.6 Å away from its original position, along [$\bar{1}$11] direction, in the midpoint of O octahedron. There is no anion present in this direction of the cell and the anion environment of O octahedron is similar to that of 3a site. So no anion disorder is observed in this case.

The calculated threshold displacement energy for Y recoil along [111] direction (Fig.5.i) is 62 eV. This value is higher than that of $Zr_{3a}$ and $Zr_{18f}$ PKA along the [111] direction. The Y PKA moves along [111] direction to form a stable cation interstitial at the bridge site between $Zr_{3a}$ and $Y_{18f}$ atoms in the nearby cell. The distance of separation of cation Frenkel pair is 5.3Å. While tracing its path, the PKA knocks out an O' atom occupied along [111] direction, which recoils along the same direction forming an anion antisite with an O atom at a distance of 5.2 Å. The recoiled O atom moves by 3.80 Å to form an anion interstitial.

Anions have lower threshold displacement energy compared to cations. At 300K, 13 eV is required for the O atom to move along [100] direction and produce a Frenkel pair separated by 2.5 Å. The cation sublattice remains unaltered. Along [110] direction the displacement energy is slightly higher (20 eV) and the Frenkel pair separation distance is 2 Å. Oxygen interstitials produced by PKAs in the [100] and [110] incident directions get stabilized in the vacant 6c site. When energy is given along [111], the O PKA which moves along the [111] direction, occupies the O' site in the nearest cell, with the movement of O' to 6c site. The process can be summarized as:

$$O_{18f}[100]\&[110] \rightarrow V_{18f} + O_{6c}$$
$$O_{18f}[111] \rightarrow O_{O'} + O'_{6c} + V_{18f}$$

O' PKA requires 17 eV to move along [100] direction and produce a Frenkel pair separated by 2.0 Å. During this process, the cation sublattice and other atoms of anion sublattice remain unaffected. In [110] and [111] directions, Frenkel pairs separated by 2.1 Å and 2.3 Å are formed.

**Defect formation energies of $Y_4Zr_3O_{12}$**

Generally, the O-D transformation in fluorite related structures is defined by two reactions: (i) antisite defects in cation sublattice and (ii) Frenkel pair formation in anion sublattice. These two defect reactions are termed as the order to disorder defect reaction pair[16]. Lower the O-D reaction pair energy, higher the radiation tolerance of the defect fluorite structure. In previous defect studies on pyrochlore structures,[38, 53, 54] the cation antisite defect is found to have the lowest formation energy and the anion Frenkel pair formation energies are significantly reduced in the presence of cation antisite defects.

The O-D reaction pair energy is considered as a good indicator of radiation tolerance in δ- phase structures also[16]. However, the defect formation energies on the $Y_4Zr_3O_{12}$ δ-phase structure is not available in any literature so far. Therefore we have attempted to study the formation of vacancies, interstitials, antisites and Frenkel pairs in $Y_4Zr_3O_{12}$ and tried to correlate O-D pair energy to the possible transformation of $Y_4Zr_3O_{12}$ to disordered fluorite structure.

The vacancy formation energies in $Y_4Zr_3O_{12}$ are calculated as: $E^X{}_{vac} = (E_{def}+E_x) - E_{total}$ where $E_{total}$ is the total energy of the supercell of the ideal δ-phase structure, $E_{def}$ is the total energy of supercell with the vacancy in it and $E_x$ is the reference energy calculated with respect to equilibrium phase of X, *i.e.*, ground state crystalline phases of Y and Zr and the gas phase of oxygen molecule. The vacancy formation energies are found to be 16.04 eV, 13.73 eV, 6.94 eV and 7.27 eV respectively for zirconium at *3a*, yttrium at *18f* and O at two sets of *18f* positions. The vacancy formation energies are comparable to that of $Y_2Ti_2O_7$ pyrochlores where 14.64 eV, 16.59 eV, 6.53 eV and 8.93 eV are required for forming Y, Ti, $O_{48f}$ and $O_{8b}$ vacancies respectively[36].

Formation energies of yttrium, zirconium and oxygen interstitials at a vacant 6c site are also considered. The interstitial defect formation energies are calculated as $E^X{}_{int} = E_{def} - (E_{total}+E_x)$, where $E_{def}$ is the energy of the supercell containing an interstitial X atom, $E_{total}$ and $E_x$ are the total energy of supercell and reference energy respectively as in the earlier case. The formation energies for interstitials are lower than those for vacancies as in the case of pyrochlores[36]. Though the 6c site is an anion vacancy, yttrium and zirconium are also found to be stable in that site. The cation interstitials are stable in the center of anion octahedron, Y being more stable than Zr by 0.1 eV. It is observed that in low energy recoil events of table.3., the displaced anion atoms tend to stabilize at the vacant 6c site and cations are stable at body center position.

Energy expenditure to form cation antisite defects, formed by replacing 18f atom with 3a atom and vice versa are calculated. The formation energy of A in B site is given by $E_B^A = E_{def} - (E_{total}+ E_A - E_B)$, where $E_{def}$ is the energy of a supercell with A in one B site, $E_A$ and $E_B$ are the energies of the supercell with A in A-site and B in B-site. Due to the similarity in cation radii, antisite pair defects are the most stable defect species in δ-phase. An additional zirconium in an *18f* yttrium site is more probable than an additional Y due to its slightly smaller radius. Likewise, the yttrium in a 3a site is less stable than the other two configurations, due to its larger ionic radius. The antisite pair defect is introduced by exchanging neighboring 18f and 3a atoms. The energy liberated or absorbed during an antisite pair formation is : $E_{AS} = E_{def} - E_{total}$, where $E_{def}$ is the energy of supercell with an antisite pair. Antisite formation energy is 0.54 eV. Antisite pair formation between *18f* cation sites is negligibly small as the structure suggests. This considerably reduces the O-D defect reaction pair energy compared to other pyrochlores.

The calculated Frenkel pair energy values show that two sets of oxygen have different vacancy formation energies and Frenkel pair formation energies due to their different chemical environments.

**Discussions**

The minimum $E_d$ values of $Y_4Zr_3O_{12}$ are found to be 28 eV for $Zr_{3a}$ PKA, 40 eV for $Zr_{18f}$ and 50 eV for Y recoils. For O and O' minimum values are 13 eV and 16 eV respectively. $Zr_{3a}$, $Zr_{18f}$ and O have the least $E_d$ values along [111] direction. Yttrium has a minimum $E_d$ along [110] direction and O' has minimum value along [100] direction. The directions except for direction in which inherent structural vacancies are aligned, cation tends to occupy another cation site. Anions tend to stabilize at vacant 6c site. The body center position, inside anion octahedron, also is a stable site for Zr cation.

Threshold displacement energies of anions are found to be dependent on the atomic environment and nature of the chemical bond [32, 55]. The O has the atomic environment of $Y_4$ tetrahedron and O' has the atomic environment of $Y_3ZrO$ trigonal bipyramid. The minimum displacement energy of O' is 3eV higher than that of O. The Zr-O distance in coordination polyhedron is smaller than Y-O distance suggesting that the Zr-O bond is more covalent in nature [39] and hence the difficulty in displacing O' compared to O.

Threshold displacement energies of $Y_2Ti_2O_7$, which is the oxide precipitate in conventional Ti containing ODS steels, has been computed by H. Y. Xiao et al[36] in detail. The minimum $E_d$ values of Y, Ti, $O_{48f}$ and $O_{8b}$ are 35.1 eV, 35.4 eV, 13 eV and 20 eV respectively. The Ed values of cations are lower than that of $Y_4Zr_3O_{12}$ computed in this paper and those of anions are similar.

According to Trachenko et al., a material with high displacement energy is expected to have higher radiation tolerance because of the smaller number of initial defects upon irradiation [55]. The irradiation resistance of GaN compared to GaAs is attributed to the higher threshold displacement energy of GaN[56]. In fluorite-structured oxides, $ThO_2$ and $CeO_2$, calculated threshold displacement energy of Th is more than that of Ce and those of oxygen are comparable (For $ThO_2$, $E_{d,Th}$ – 53.5 eV, $E_{d,O}$ = 17.5 and for $CeO_2$ , $E_{d,Ce}$=46 eV and $E_{d,O}$ = 20 eV) [33]. The experimental evidence shows that defect accumulation and radiation damage are more in $CeO_2$[57].

According to Xiao et al[39] and Wang et al[37], in pyrochlores, the higher the threshold displacement energy, higher the energy barrier for O- D transformation. So the pyrochlores with higher displacement energy tend to amorphize faster than those with lower displacement threshold. Experimentally, $Gd_2Zr_2O_7$ is the most radiation resistant zirconate pyrochlore. But the ab initio molecular dynamics simulations give higher values for threshold displacement energies ($E_{d,Gd}$ = 31 eV, $E_{d,Zr}$ = 39 eV, $E_{d,O8b}$=13 eV, $E_{d,O48f}$= 9.5eV) [58] compared to $La_2Zr_2O_7$ ($E_{d,La}$= 29.5 eV, $E_{d,Zr}$= 39.5 eV, $E_{d,O8b}$= 10.5 eV $E_{d,O48f}$ =15.5 eV) ,$Nd_2Zr_2O_7$ ($E_{d,Nd}$=21.5 eV, $E_{d,Zr}$=27 eV, $E_{d,O48f}$ = 6.5 eV, $E_{d,O8b}$ = 3.0 eV) and $Sm_2Zr_2O_7$ ($E_{d,Nd}$= 21.5 eV, $E_{d,Zr}$= 26.5 eV, $E_{d,O48f}$ =7.0 eV, $E_{d,O8b}$= 3.0 eV), which are comparatively less radiation resistant[39].So the argument of higher radiation resistance for lower threshold displacement energy is not valid in these cases. Therefore, further theoretical and experimental studies are required to link threshold displacement energy and radiation stability of material.

The defect formation energies are compared to previously calculated values of pyrochlore structures. The cation antisite pair formation energy (0.54 eV for 3a -18f cation swap) lower than that of all the pyrochlores data available so far (1.77[59] $Y_2Ti_2O_7$), 1.97[59] ($Y_2Sn_2O_7$), 1.9 [60]($Gd_2Ti_2O_7$) 1.8[60]($Gd_2Zr_2O_7$)) and anion Frenkel pair formation energies are less than $Y_2Ti_2O_7$ but more than $Gd_2Zr_2O_7$. The O-D reaction pair energy is 1.66 eV, which is lower than all the data on pyrochlores [16]. So, the defect formation energy values suggest rapid transformation to disordered fluorite structure and hence more resistance to amorphization for $Y_4Zr_3O_{12}$ compared to other pyrochlore structures. These results convey that Zr/Al ODS alloys which have better corrosion resistance properties compared to the conventional Ti-ODS alloys may also possess superior radiation resistance.

**Conclusions**

Low energy recoil events and defect formation energies in $Y_4Zr_3O_{12}$ have been investigated using ab initio methods based on density functional theory. Our density functional theory calculations show that among available 18f sites in the δ-phase structure, the Zr atoms in $Y_4Zr_3O_{12}$ prefer to occupy the farthest corners of cation octahedron. The threshold displacement energies are found to be anisotropic, strongly dependant on the incident direction. The minimum threshold displacement energies of $Zr_{3a}$, $Zr_{18f}$, and $Y_{18f}$ are 28 eV, 40 eV, and 50 eV respectively. Anions have smaller threshold displacement energies than cations and defect generation mechanisms of cations involve more number of defects, suggesting that anion disorder is more probable than cation disorder. The Y – $Zr_{18f}$ antisite formation energy has the least value among defect formation energies considered. So upon irradiation, the 18f atoms can interchange position and continue to be in the δ-phase structure before transforming into the defect fluorite phase.

**Table.1.** Formation energies/unitcell for different configurations of Y and Zr in 18f positions in $Y_4Zr_3O_{12}$ structure.

| Configuration | | Formation energy/unit cell (eV) |
|---|---|---|
| YZ1 | 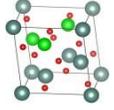 | -50.80 |
| YZ2 | 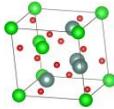 | -51.18 |
| YZ3 | 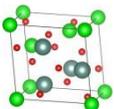 | -51.21 |
| YZ4 | 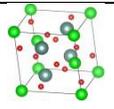 | -51.36 |
| YZ5 | 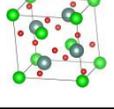 | -51.36 |
| YZ6 | 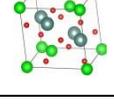 | -51.36 |

**Table.2.** The formation energies of different structures considered in Y-Zr-O system.

| Structure | O vacancy Environment | Space Group | Formation energy/atom (eV) |
|---|---|---|---|
| $Y_2Zr_2O_7$ - 1 | 4Zr | $Fm\bar{3}m$ | -2.28 |
| $Y_2Zr_2O_7$ - 2 | 4Y | $Fm\bar{3}m$ | -2.27 |
| SQS | - | - | -2.23 |
| δ-phase | 2Zr2Y | $R\bar{3}$ | -2.69 |

**Table.3.** The threshold displacement energies of Y, Zr and O in $Y_4Zr_3O_{12}$ structure. ($A_B$ is A atom occupying site of B atom.)

| Direction | $E_d$(eV) | Defect Configurations | Separation between defects (Å) |
|---|---|---|---|
| $Zr_{3a}$[100] | 65 | $Zr_Y$ | 3.6 |
| | | $O'_{int} + O'_{vac}$ | 3.7 |
| $Zr_{3a}$[110] | 60 | $Zr_Y$ | 5.0 |
| | | $O_{O'}$ | 3.0 |
| | | $O'_{int} + O_{vac}$ | 2.8 |
| $Zr_{3a}$[111] | 28 | $Zr_{3a\_int} + Zr_{3a\_vac}$ | 4.7 |
| $Zr_{18f}$[100] | 65 | $Zr_{Zr}$ | 4.6 |
| | | $Zr_{int}$ | 2.4 |
| | | $O'_{int} + O'_{vac}$ | 2.4 |
| | | $O'_{int} + O'_{vac}$ | 2.8 |
| $Zr_{18f}$[110] | 69 | $Zr_{Zr}$ | 5.3 |
| | | $O_{O'}$ | 3.38 |
| | | $O'_{O'}$ | 3.56 |
| | | $O_{vac} + O'_{int}$ | 2.8 |
| | | $Zr_{int}$ | 3.41 |
| $Zr_{18f}$[111] | 40 | $Zr_{int} + Zr_{18f\_vac}$ | 5.12 |
| | | $O'_{O'}$ | 4.4 |
| | | $O'_{vac} + O'_{int}$ | 3.3 |
| | | $O_{vac} + O_{int}$ | 2.8 |
| Y[100] | 65 | $Y_Y$ | 3.9 |
| | | $O_{vac} + O_{int}$ | 3.2 |
| | | $O'_{vac} + O'_{int}$ | 2.6 |
| Y[110] | 50 | $Y_{3a}$ | 5.4 |
| | | $Y_{vac} + Zr_{int}$ | |
| Y[111] | 62 | $Y_{int} + Y_{vac}$ | 5.2 |
| | | $O_{int} + O_{vac}$ | 5.3 |
| | | $O_{int} + O_{vac}$ | 3.8 |
| O [100] | 13 | $O_{int} + O_{vac}$ | 3.2 |
| O[110] | 20 | $O_{int}+O_{vac}$ | 2.35 |
| O[111] | 13 | $O_{int}+O_{vac}+O'_{int}+O'_{vac}$ | 2.0 |
| O'[100] | 16 | $O'_{int}+O'_{vac}$ | 2.0 |
| O'[110] | 22 | $O'_{int}+O'_{vac}$ | 2.1 |
| O'[111] | 26 | $O'_{int}+O'_{vac}$ | 2.3 |

**Table.4.** Defect formation energies in $Y_4Zr_3O_{12}$ structure.

| Species | Formation Energy (eV) |
|---|---|
| **Vacancy** | |
| O | 6.94 |
| Y | 13.72 |
| Zr | 16.04 |
| O' | 7.27 |
| **Interstitial** | |
| O_6c | 2.85 |
| Y_6c | 4.54 |
| Zr_6c | 4.29 |
| Y_center | 3.75 |
| Zr_center | 3.65 |
| O_6c+ O_6c | 0.64 |
| **Antisite** | |
| Y_3a | 2.76 |
| Zr_18f | 2.03 |
| Y_18f | 2.50 |
| AS$_{pair}$ | 0.54 |
| **Frenkel Pairs** | |
| O_6c+O_vac | 1.12 |
| Zr_6c+Zr3a_vac | 7.76 |
| Y_6c+Y_vac | 7.37 |
| O_6c+O'_vac | 3.21 |

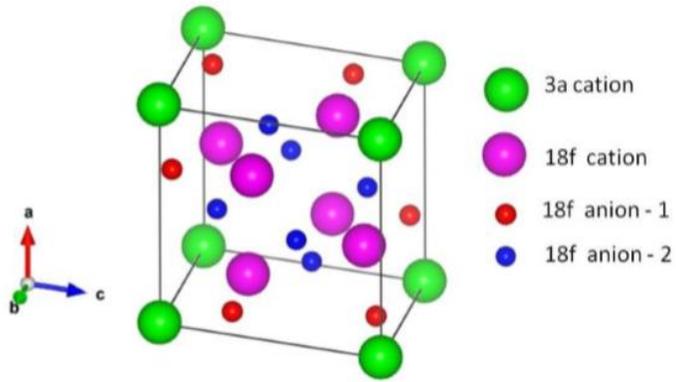

**Fig.1.** The Y$_4$Zr$_3$O$_{12}$ unitcell

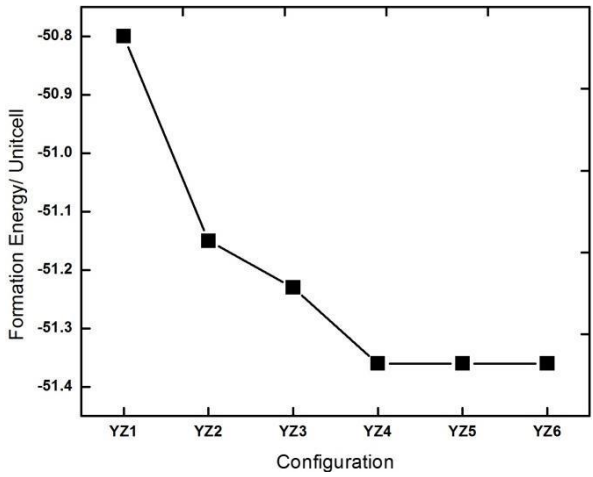

**Fig.2.** Formation energies/unitcell for different configurations of Y and Zr in 18f positions in the $Y_4Zr_3O_{12}$ structure.

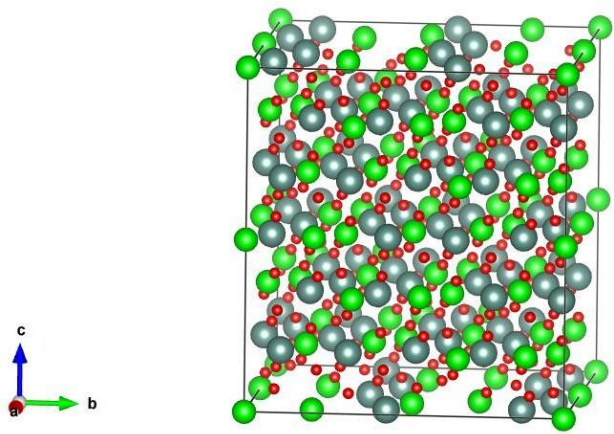

**Fig.3**. Transformed orthogonal cell of the $Y_4Zr_3O_{12}$, containing 456 atoms. a= 19.44Å b = 16.84 Å, c= 18.18 Å, α=β=γ=90°. This cell is used for AIMD calculations.

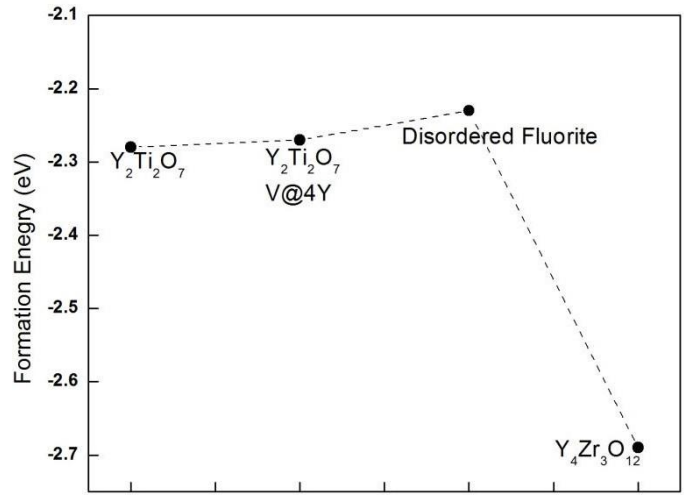

**Fig.4.** The formation energies of different structures considered in Y-Zr-O system

(a) 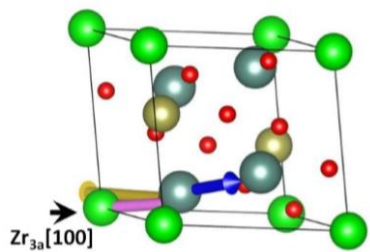

(b) 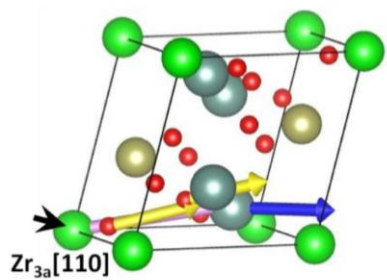

(c) 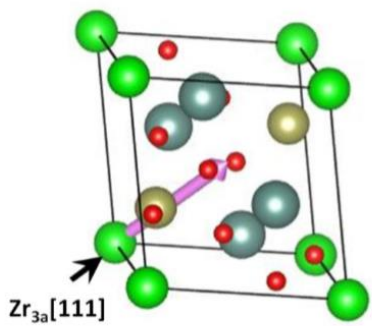

(d) 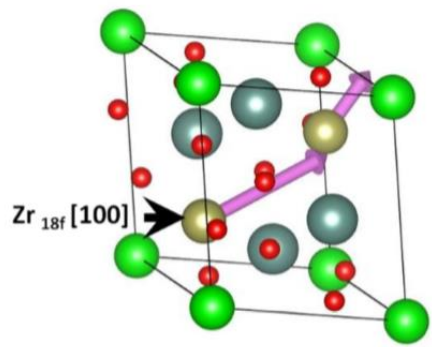

(e) 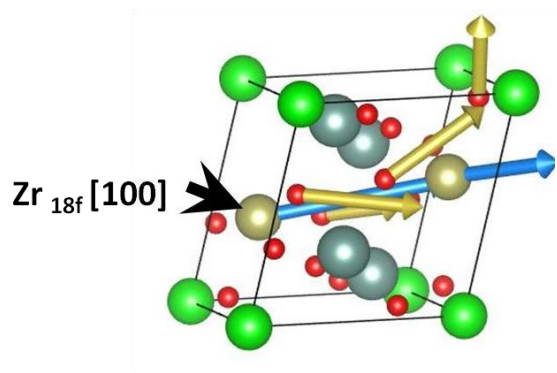

(f) 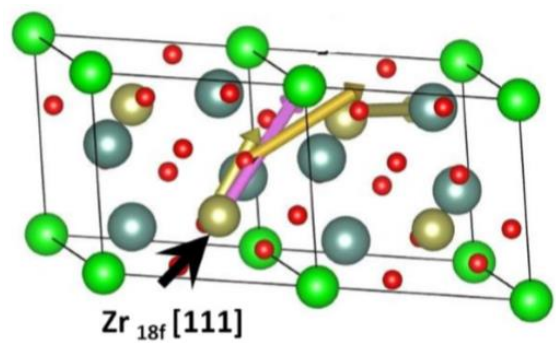

(g)

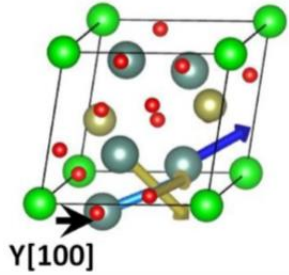

Y[100]

(h)

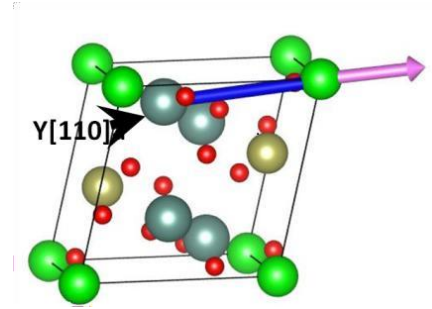

Y[110]

(i)

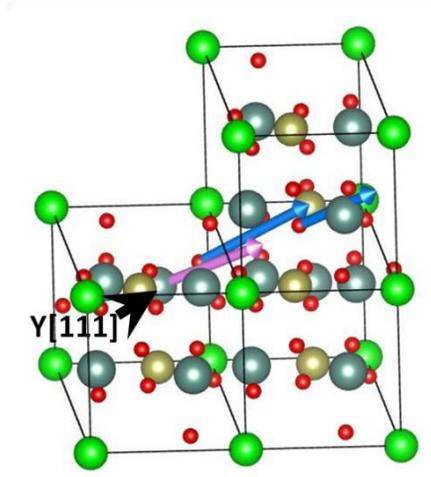

Y[111]

**Fig.5.** Defect configurations at 300K when $Zr_{3a}$, $Zr_{18f}$ and Y atoms are given energy along different directions. The direction in which energy is given is marked by a black arrow. The Purple arrow indicates Zr displacement direction, the blue arrow indicates Y displacement direction and the yellow arrow indicates oxygen displacement direction. Here, the green atoms are $Zr_{3a}$, the yellow atoms are $Zr_{18f}$, the blue atoms are $Y_{18f}$, and the red atoms are oxygen.

(a) $Zr_{3a}[100] \rightarrow Zr_Y + Y_{int} + Zr_{3a\_vac} + O_{int} + O_{vac}$

(b) $Zr_{3a}[110] \rightarrow Zr_Y + Y_{int} + Zr_{3a\_vac} + O_{O'} + O'_{int} + O_{vac}$

(c) $Zr_{3a}[111] \rightarrow Zr_{int} + Zr_{3a\_vac}$

(d) $Zr_{18f}[100] \rightarrow Zr_{Zr} + Zr_{int} + Zr_{18f\_vac} + 2O'_{int} + 2O'_{vac}$

(e) $Zr_{18f}[110] \rightarrow Zr_{Zr} + Zr_{int} + Zr_{18f\_vac} + O'_O + 2O_{int} + O_{vac} + O'_{vac}$

(f) $Zr_{18f}[111] \rightarrow Zr_{int} + Zr_{vac} + O'_{O'} + 2O'_{int} + 2O'_{vac}$

(g) $Y[100] \rightarrow Y_Y + Y_{int} + Y_{vac} + O_{int} + O'_{int} + O_{vac} + O'_{vac}$

(h) $Y[110] \rightarrow Y_{Zr\_3a} + Zr_{int} + Y_{vac}$

(i) $Y[111] \rightarrow Y_Y + Y_{int} + Y_{vac} + O_{int} + O'_{int} + O_{vac} + O'_{vac}$

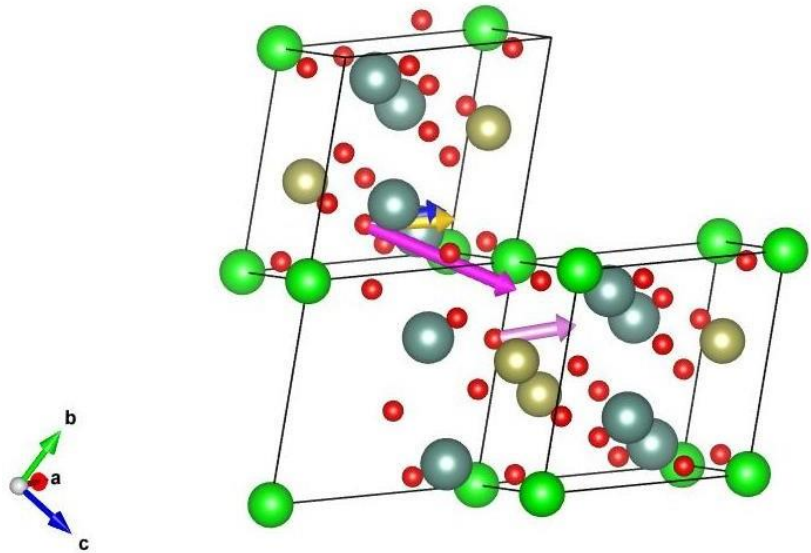

**Fig.6.** The displacements of O atom when incident energy is given along [100] (Blue arrow), [110] (Yellow arrow) and [111] (Purple arrow) directions. The first two results in FP with interstitial occupied in 6c site and third one results in an anion antisite pair and FP with interstitial in 6c site. Here green atoms are $Zr_{3a}$, Yellow atoms are $Zr_{18f}$, blue atoms are $Y_{18f}$ and red atoms are oxygen.

**Data availability statement**
The data that support the findings of this study are available from the corresponding author upon reasonable request.